\documentclass{article}
\usepackage[T1]{fontenc}
\usepackage[utf8]{inputenc}
\usepackage{ismir} %
\usepackage{amsmath,cite,url}
\usepackage{graphicx}
\usepackage{tabularx} 
\usepackage{color}
\usepackage{amssymb}
\usepackage{subcaption}
\usepackage{booktabs}
\usepackage{multirow}
\usepackage{makecell}
\usepackage{float}
\usepackage{siunitx}
\usepackage[hidelinks,unicode]{hyperref}

\newcolumntype{C}{>{\centering\arraybackslash}X}
\newcolumntype{P}[1]{
  S[
    table-format=#1,
    table-space-text-post={\%},
    reset-text-series=false,
    mode=text
  ]<{\,\%}
}
\newcommand{\mc}[1]{\multicolumn{1}{c}{#1}}
\newcommand{\mcc}[1]{\multicolumn{1}{c|}{#1}}
\title{BARS: Beat-Adaptive Rap Synthesis via Flow Matching}

\multauthor
  {Taemin Cho \hspace{1cm} Hector Martel \hspace{1cm} Enyan Koh}
  {BandLab Technologies, Singapore\\
  {\tt\small taemin.cho@bandlab.com}
  }
\def\authorname{T. Cho, H. Martel, and E. Koh}

\begin{document}

\maketitle

\begin{abstract}
We present BARS (Beat-Adaptive Rap Synthesis), a non-autoregressive rapping voice generation model based on Flow Matching, conditioned on a beat track (instrumental backing), lyrics, and a short reference voice for zero-shot timbre control. Unlike prior autoregressive rap synthesis, where rap pace and output duration are strictly determined by the beat, BARS enables independent control over the rap tempo, allowing flexible-length generation without explicit duration prediction or phoneme-level alignment.
We encode lyrics using a pretrained ByT5 encoder with a lightweight Transformer adapter trained jointly with the model, without relying on speech-pretrained text encoders, enabling natural handling of non-standard orthography (e.g., slang, abbreviations) and numerals common in rap lyrics.
To address alignment instabilities, such as token skipping or repetition, under strong beat conditioning, we propose Asymmetric Scaled RoPE for cross-attention, which introduces a positional structure to facilitate sequential lyric-audio alignment learning. Furthermore, the model supports partial re-generation via span masking, enabling lyric-level editing while preserving the surrounding vocal style. 
Experiments demonstrate that the proposed model achieves high lyric fidelity, natural rap flow that is rhythmically coherent with the beat, and robust tempo control across a wide range of rapping styles and speeds.
\end{abstract}

\section{Introduction}\label{sec:introduction}

Rap music occupies a unique intersection of rhythm, language, and vocal expression, where the vocal delivery is governed by a rigid rhythmic contract with the underlying beat. Unlike melodic singing, rapping requires syllables to fall precisely within specific metrical windows while preserving the semantic integrity of the lyrics~\cite{ohriner2019flow}. This dual constraint of temporal synchronization and semantic fidelity makes rapping voice generation significantly more challenging than conventional Text-to-Speech (TTS) or singing voice synthesis.

Existing Singing Voice Synthesis (SVS) systems~\cite{chen2020hifisinger, liu2022diffsinger, zhang2023wesinger} typically require predefined pitch and duration inputs, which are fundamentally incompatible with the nature of rap. 
Indeed, rap delivery is largely pitch-agnostic and characterized by irregular rhythmic variations.
The essence of rap performance, commonly referred to as \textit{flow}~\cite{ohriner2019flow}, 
emerges from dynamic stress patterns, syncopation, and deliberate timing deviations that cannot be captured by fixed pitch and duration templates.
Recent work on beat-conditioned rap generation, notably Freestyler~\cite{ning2025drop}, has made meaningful progress in this direction. 
However, its autoregressive nature results in slow sequential decoding, and both the rapping pace and output duration are solely determined by the beat track, leaving the model with no independent control over vocal delivery. 

In this work, we propose BARS (Beat-Adaptive Rap Synthesis)\footnote{Demo: \url{https://bars-ismir2026.github.io/bars}}, a non-autoregressive rapping voice generation model built on Flow Matching~\cite{lipman2023flow}.
The model is conditioned on a beat track, lyrics, and a short reference voice, and generates the corresponding rap vocal track, which can subsequently be mixed with the input beat track to produce the final output. It is designed as a creative tool to help rap artists prototype and refine their vocal delivery.
Its non-autoregressive design enables independent control over the tempo of the generated vocals, allowing artists to explore diverse rhythmic interpretations of the same lyrics over a given beat. In addition, training with span masking allows partial re-generation, making it possible to revise specific phrases while preserving the vocal style and rhythmic consistency of the surrounding performance.
The model further supports zero-shot timbre control, enabling synthesis in the voice of an unseen speaker using only a short reference voice sample.

For lyric encoding, we leverage a pretrained ByT5 encoder~\cite{xue2022byt5}, which provides rich byte-level linguistic representations learned from large-scale text data. 
To map these embeddings for cross-modal alignment with audio latents, we introduce a lightweight Transformer-based adapter.
We further observe that strong beat conditioning can disrupt positional tracking over the lyric sequence during cross-attention, leading to skipped or repeated lyric tokens. To mitigate this, we propose an \textit{Asymmetric Scaled RoPE} (A-RoPE) mechanism that applies standard RoPE to the audio query side while applying length-proportionally scaled RoPE to the lyric key side, encouraging more consistent sequential coverage of the lyric tokens. 

To the best of our knowledge, this is the first non-autoregressive Flow Matching-based framework for rapping voice generation that achieves flexible-length output without relying on explicit duration prediction, phoneme-level alignment, or speech-specific encoders.

\section{Related Work}

\subsection{Rap Generation}

Previous works on rap generation have largely focused on the linguistic dimensions of the genre. Early systems such as DopeLearning~\cite{malmi2016dopelearning} and GhostWriter~\cite{potash2015ghostwriter} addressed lyric composition through information retrieval using RankSVM and LSTM-based sequence modeling respectively, prioritizing rhyme quality and stylistic consistency. Rapformer~\cite{nikolov2020rapformer} further extended this direction with a Transformer-based denoising autoencoder capable of transferring non-rap text into rap lyrics. DeepRapper~\cite{xue2021deeprapper} marked a step toward rhythmic awareness by explicitly modeling rhyme and beat alignment through a reverse-order language model with inserted beat tokens. Despite this progress, these systems operate entirely in the text domain and produce no vocal audio, leaving the acoustic realization of rap performance largely unaddressed.

The shift toward audio-based rap generation is more recent. Freestyler~\cite{ning2025drop} represents the first system to synthesize rapping vocals directly from lyrics and accompaniment, employing a three-stage hierarchical framework to achieve stylistic and rhythmic alignment with the backing track. 
However, its autoregressive architecture introduces certain limitations. In particular, the model operates in a unidirectional manner, which restricts its ability to consider the global structure of the accompaniment when planning vocal delivery. As the model is trained to synchronize closely with local, frame-level acoustic features, it has limited capacity to adjust delivery speed based on the overall duration of the beat. Consequently, the generated vocals tend to follow the input track's rhythm, with limited flexibility in pacing control across varying track lengths.
These limitations result in early termination when the beat is too short and overly extended outputs when it is excessively long. Although Freestyler incorporates a random masking strategy during training to alleviate these issues, explicit control over pacing independent of the beat track remains limited. Additionally, as model weights are not publicly available, direct quantitative comparison is not feasible.

\subsection{Non-Autoregressive Speech Synthesis}

Non-autoregressive approaches to TTS have advanced rapidly through diffusion-based frameworks. 
E3 TTS~\cite{gao2023e3} explores end-to-end diffusion-based TTS without explicit alignment supervision, but its U-Net backbone constrains generation to fixed-length outputs. NaturalSpeech~\cite{shen2024naturalspeech, ju2024naturalspeech} and Voicebox~\cite{le2023voicebox} overcome this limitation by adopting Transformer architectures that support variable-length generation. However, they depend on phoneme-level alignment and separate duration prediction modules, which introduce additional training complexity.

E2 TTS~\cite{eskimez2024e2} takes a further step toward simplicity by eliminating all explicit alignment constraints. By framing TTS as an audio infilling task and using filler tokens to reconcile text and audio lengths, E2 TTS learns alignment implicitly during training. DiTTo-TTS~\cite{lee2025dittotts} similarly avoids explicit duration supervision via cross-attention between text and audio representations. However, it reports that ByT5, a text-only pretrained encoder, substantially underperforms speech-pretrained encoders such as SpeechT5~\cite{ao2022speecht5} when codec-level text-audio fine-tuning is absent. This finding suggests that at least one side of the text-audio interface must encode cross-modal information to effectively bridge the modality gap.

\section{Method}

Our approach builds on Flow Matching~\cite{lipman2023flow}, ByT5~\cite{xue2022byt5}, RoPE~\cite{su2024roformer}, and an audio VAE~\cite{evans2024longform}; our contributions lie specifically in the cross-modal adapter (\autoref{sec:input_ref}) and the A-RoPE mechanism (\autoref{sec:rope}).

\subsection{Input Representations}\label{sec:input_ref}
The model takes three inputs: a beat track, lyrics, and a 3-second reference voice. The beat track and reference voice, along with the ground truth rap vocal during training, are encoded to compact latent sequences using an audio VAE trained following the architecture of Stable Audio 2.0~\cite{evans2024longform} on proprietary music data comprising diverse music genres, sound effects, and vocal recordings.

Lyrics are encoded using a pretrained ByT5-base encoder\footnote{\url{https://huggingface.co/google/byt5-base}}. 
Rather than fine-tuning the audio encoder or relying on a speech-pretrained text encoder, we introduce a lightweight Transformer encoder~\cite{vaswani2017attention} as a cross-modal adapter on the text side. This adapter is trained jointly with the Flow Matching network and learns to map ByT5 embeddings into a representation space compatible with audio latents for cross-attention. This design preserves the byte-level coverage of ByT5, which can naturally handle numbers, symbols, and non-standard orthography common in rap lyrics, while avoiding the limitations of speech-pretrained encoders such as SpeechT5, which relies on a fixed phonetic vocabulary that treats such inputs as out-of-vocabulary tokens. 

\subsection{Model Input Construction}

Inspired by \cite{lee2025dittotts}, we incorporate span masking into training to enable lyric infilling.
While DiTTo-TTS~\cite{lee2025dittotts} leverages this objective to naturally acquire zero-shot voice cloning capability, our setting differs: since the generated vocal must cover the full lyric sequence from the start of the beat, we instead provide speaker identity via a 3-second reference voice prepended to the input sequence, enabling zero-shot timbre control without requiring a paired transcript. 

Let $z_1\in \mathbb{R}^{D \times N}$ denote the ground truth rap vocal latent sequence, where  $D$ is the latent dimension and $N$ is the sequence length. A binary mask $m \in \{0,1\}^N$ is constructed by randomly selecting a contiguous span. Following \cite{lee2025dittotts}, the mask either covers the entire sequence with probability 0.1, or masks a random contiguous segment spanning 70\% to 100\% of the sequence length. The masked input is formed as: 
\begin{equation}
\tilde{z}_t = m \odot z_t + (1 - m) \odot z_1
\end{equation}
Here, $\odot$ denotes element-wise multiplication, $z_t = (1-t)\cdot z_0 + t \cdot z_1$ is the Flow Matching interpolant applied to masked positions, where $z_0 \sim \mathcal{N}(0,\mathbf{I})$ is Gaussian noise, and $t \in [0, 1]$ is the timestep sampled uniformly during training. The unmasked positions (i.e., where $m=0$) retain the ground truth values $z_1$, conditioning the model on the surrounding context.
The training loss is computed only over the masked regions:
\begin{equation} 
\mathcal{L} = \mathbb{E}_{t, z_0} \left\| m \odot (u_t - v_\theta(\tilde{z}_t, t, c)) \right\|^2
\end{equation}
where $u_t = z_1 -z_0$ is the Flow Matching target, and $c$ represents the conditioning inputs.

\begin{figure}[t] 
  \centering
  \includegraphics[width=0.75\linewidth]{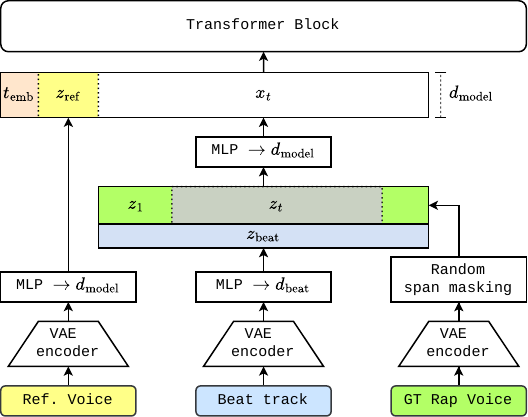}
  \caption{
  Input construction: the rap vocal, beat track, and reference voice are encoded via a shared VAE encoder. A span mask is applied to the vocal latent, and masked regions are replaced with the Flow Matching interpolant $z_t$. The beat latent is concatenated along the channel dimension and projected to form $x_t$, which is prepended with $z_\text{ref}$ and timestep embedding $t_\text{emb}$.
  }
  \label{fig:audio_input} 
\end{figure}

The beat track latent is projected to a lower-dimensional representation $z_\text{beat} \in \mathbb{R}^{d_\text{beat} \times N}$ and concatenated with $\tilde z_t$ along the channel dimension, providing a frame-level rhythmic scaffold. The concatenated sequence is then projected to the model dimension $d_\text{model}$ to produce the audio input tokens $x_t$.
As mentioned above, the reference voice latent is projected to $d_\text{model}$ and prepended to the audio input tokens, along with a sinusoidal embedding of the timestep $t$, following Stable Audio 2.0~\cite{evans2024longform}. The full input sequence to the Transformer is thus:
\begin{equation}
\text{input} = [t_\text{emb},\, z_\text{ref},\, x_t]
\end{equation}
This process is illustrated in \autoref{fig:audio_input}.

\subsection{Model Architecture}

\begin{figure}[t] 
  \centering
  \includegraphics[width=0.98\linewidth]{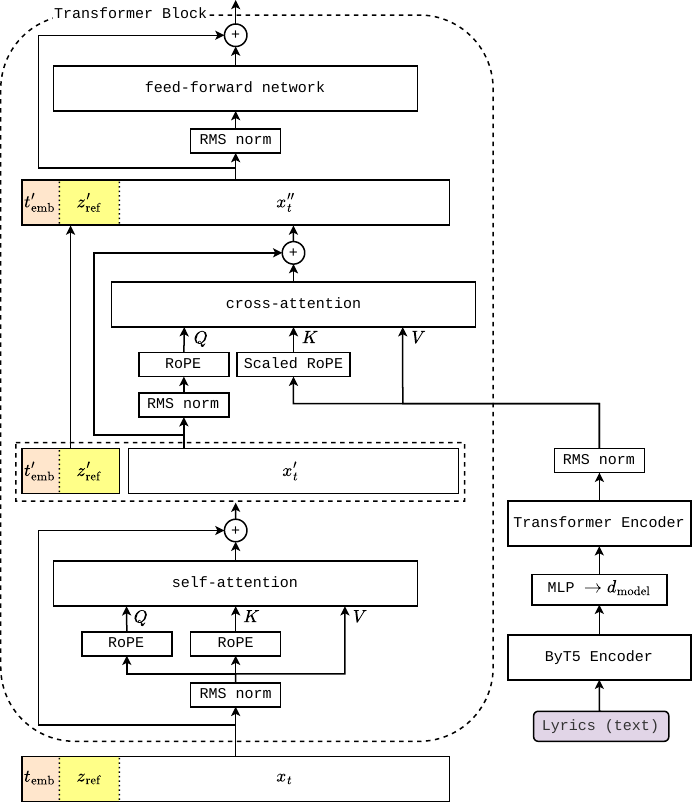}
  \caption{
  Architecture of a single Transformer block: RoPE is applied in self-attention with position indices reset per segment. Prior to cross-attention, conditioning tokens are removed and only audio latent tokens serve as queries, with standard RoPE on the audio side and length-scaled RoPE on the lyric key side. The conditioning tokens are reintroduced before the feed-forward network.
  }
  \label{fig:transformer_block} 
\end{figure}

The input sequence is processed by a stack of Transformer blocks, as illustrated in \autoref{fig:transformer_block}. Each block consists of a bidirectional self-attention layer, a cross-attention layer, and a feed-forward network. In the self-attention layer, Rotary Position Embeddings (RoPE)~\cite{su2024roformer} are applied to both the query and key, with position indices reset to 0 at the start of each segment (i.e., the prepended conditioning tokens and the audio latent tokens are each indexed independently starting from 0). This position-decoupled scheme allows the timestep and reference to interact with the audio latents as global context without introducing positional bias.
 
Lyric representations are injected into the Transformer blocks via cross-attention layers, enabling alignment between audio and text without explicit duration prediction or phoneme-level supervision.
Prior to cross-attention, the prepended conditioning tokens ($t_\text{emb}'$ and $z_\text{ref}'$) are temporarily removed from the sequence and only the audio latent tokens $x_t'$ are used as queries, while the lyric representations serve as keys and values. 
Since the timestep and reference function as global conditioning signals rather than content tokens that align with lyrics, they are excluded from the cross-attention query. This is possible because their information is already propagated to the audio latent tokens through the preceding self-attention.
This design ensures a positionally coherent application of A-RoPE, described in \autoref{sec:rope}. Finally, the conditioning tokens are reintroduced before the feed-forward network.

\subsection{Asymmetric Scaled RoPE for Cross-Attention}\label{sec:rope}

\begin{figure}[t]
    \centering
    \begin{subfigure}[b]{0.495\columnwidth}
        \centering
        \includegraphics[width=\linewidth]{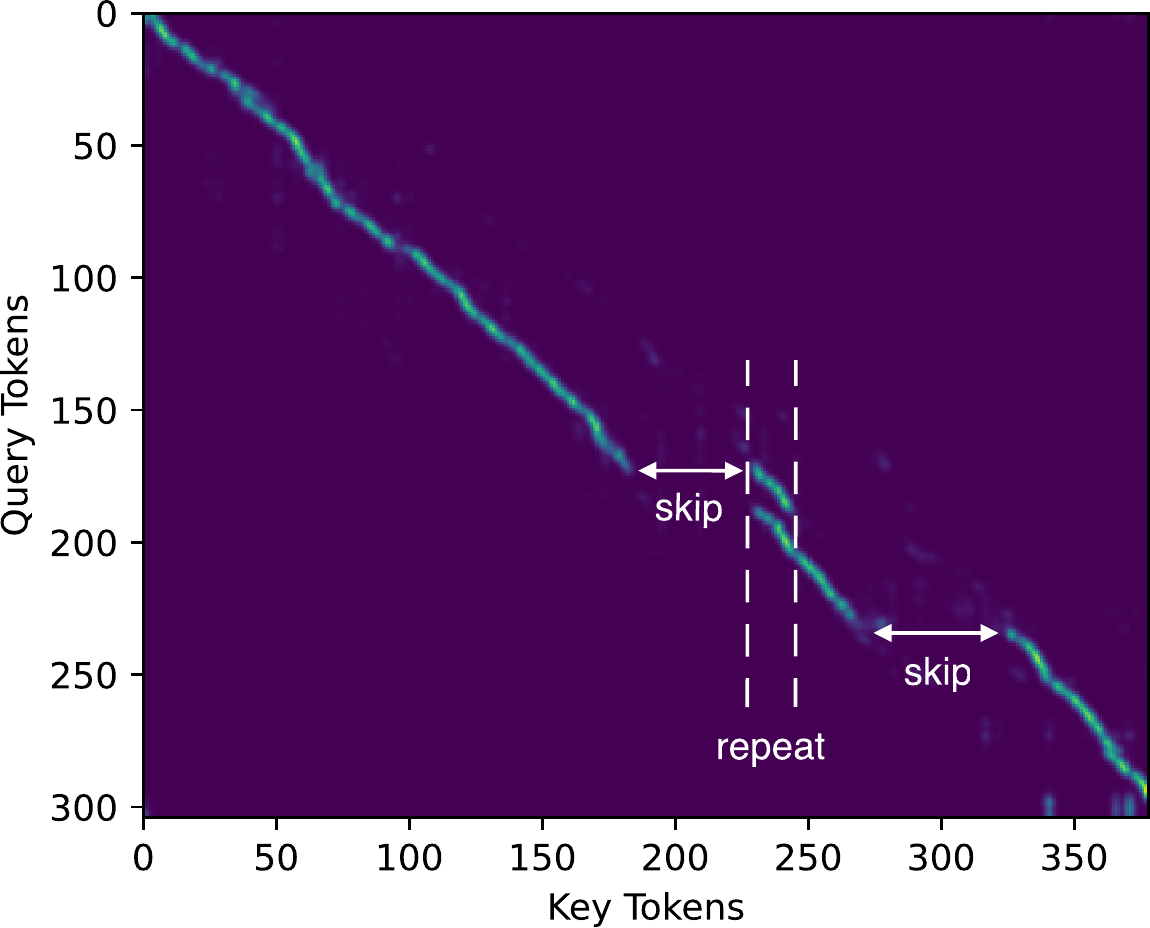}
        \caption{w/o A-RoPE}
    \end{subfigure}
    \hfill
    \begin{subfigure}[b]{0.495\columnwidth}
        \centering
        \includegraphics[width=\linewidth]{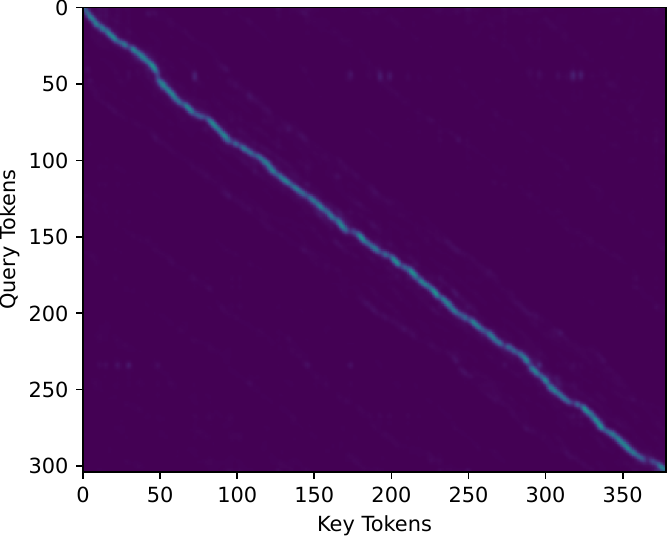}
        \caption{w/ A-RoPE}
    \end{subfigure}
    \caption{Cross-attention maps (mid-layer). (a) Abrupt horizontal shifts indicate skipping, while the double-line pattern denotes repetition via redundant attention. (b) Attention follows a smooth diagonal trajectory.}
    \label{fig:attn_map}
\end{figure}

As previously noted, without positional guidance in cross-attention, the model tends to prioritize rhythmic alignment with the beat at the expense of accurate lyric reproduction.
We observe that this behavior correlates with the strength of the beat conditioning: as the beat input is scaled down or removed, lyric accuracy improves, albeit at the cost of rhythmic alignment. The issue becomes more pronounced when the lyric length deviates substantially from what the beat can naturally accommodate. To address this, we introduce an A-RoPE mechanism in the cross-attention layers. 
 
In standard RoPE, the relative position between query position $i$ and key position $j$ is encoded through the rotation angle difference, which biases attention toward nearby positions. In our cross-attention, the query (audio latent) and key (lyric embedding) sequences differ in length: the audio sequence has length $N$ while the lyric sequence has length $L$. 
We apply standard RoPE to the audio query side, while scaling the RoPE position indices on the lyric key side by a factor of $S = N / L$. Specifically, for a lyric key at position $j$, its RoPE rotation angle is computed using the scaled position $j \cdot S$ instead of $j$.
Intuitively, this stretches the lyric position indices to span the temporal extent of the audio as if a lyric sheet were laid out along the track's timeline, encouraging the model to learn a sequential correspondence between audio frames and lyric tokens: audio frame $i$ is biased toward lyric token $j \approx i \cdot L / N$.
This mechanism allows the model to deviate from strict diagonal alignment where appropriate (e.g., held syllables or rhythmic pauses) while discouraging the skip and repeat patterns that arise in the unconstrained setting.
\autoref{fig:attn_map} visualizes cross-attention maps from models with and without A-RoPE. Without it, the attention pattern exhibits frequent skipping and repetition, whereas the model with A-RoPE maintains a coherent diagonal trajectory without skipping or repetition.

Closely related to our approach, LARoPE~\cite{kim2026length} proposes a symmetric length-normalization of RoPE for TTS cross-attention, where both query and key positions are divided by their respective sequence lengths. Our approach instead scales only the key (lyric) side by the length ratio $S = N/L$, leaving the query (audio) side as standard RoPE, which avoids introducing additional hyperparameters and is motivated by the distinct challenge of beat-conditioned generation, where rhythmic conditioning competes with lyric alignment.

\section{Experiment}

\subsection{Dataset and Implementation Details}
 
We train on approximately 1,500 hours of English rap music from both male and female artists, covering a diverse range of subgenres. Each training sample consists of a rap vocal segment with its corresponding lyrics and beat track, along with a 3-second reference clip from a different segment of the same rapper. We split the data into train/validation/test sets (90/5/5), ensuring no rapper overlap across splits.
Our model ($\sim$200M parameters) is trained for 500k steps on 8 NVIDIA H100 GPUs using AdamW (peak LR $1 \times 10^{-4}$, OneCycleLR with 2k warmup, batch size 64) with EMA (decay 0.9999). While the model architecture imposes no inherent length constraint, training samples are capped at 60 seconds for memory efficiency; we have not evaluated inference beyond this duration.
At inference, the output length is set to that of the given beat track, and we integrate the flow ODE using the Euler method (25 steps) with classifier-free guidance (scale 3.0). On a single NVIDIA RTX 4090 GPU, our model achieves a real-time factor (RTF) of 0.027 (10 seconds of audio in 0.27 seconds), including VAE encoding/decoding and text encoding.
 
\subsection{Evaluation Metrics}

We evaluate generated rap vocals using the following metrics. \textit{Word Error Rate (WER)} is computed by transcribing generated vocals with Whisper large-v3~\cite{radford2023robust} and comparing the resulting transcripts against the ground-truth lyrics. We further decompose WER into deletion, insertion, and substitution rates to distinguish lyric skipping, repetition, and mispronunciation. As rap vocals pose inherent challenges for ASR systems, WER should be interpreted relative to the GT transcription error (5.1\% in \autoref{tab:ablation}) rather than as an absolute measure of generation quality. 
\textit{Speaker similarity} is measured as the cosine similarity between speaker embeddings extracted from the reference and generated audio using WavLM-Base+~\cite{chen2022wavlm}, averaging over non-overlapping 3-second windows for vocals longer than the reference. 
\textit{DNSMOS}~\cite{reddy2021dnsmos} is included as a supplementary automatic proxy for audio quality. Perceptual quality is primarily assessed through a subjective MOS study on a 5-point Likert scale, evaluating rhythmic flow (MOS-F), speaker similarity (MOS-S), and audio quality (MOS-Q), with details provided in \autoref{sec:human_mos}.

\subsection{Ablation Study}

\begin{table}[t]
  \centering
  \small
    \begin{tabularx}{\columnwidth}{@{\hspace{0pt}}C@{\hspace{3pt}}C@{\hspace{3pt}}C@{\hspace{3pt}}|@{\hspace{0pt}}P{1.1}@{\hspace{0pt}}P{1.1}@{\hspace{0pt}}P{1.1}@{\hspace{0pt}}P{1.1}@{\hspace{0pt}}}
    \toprule
    Encoder & Adapter &\mbox{A-RoPE} & \mc{WER\hspace{0.1em}$\downarrow$} & \mc{Del\hspace{0.1em}$\downarrow$} & \mc{Ins\hspace{0.1em}$\downarrow$} & \mc{Sub\hspace{0.1em}$\downarrow$} \\
    \midrule 
    GT & \multicolumn{2}{c|}{---} & 5.1 & 0.7 & 0.6 & 3.8 \\
    \midrule
    \multirow{4}{*}{ByT5}
    & \checkmark & \checkmark & \bfseries 5.1 & \bfseries 1.1 & \bfseries 1.1 & \bfseries 2.9 \\    
    & \checkmark &            & 6.0 & 1.5 & 1.2 & 3.3 \\
    &            & \checkmark & 9.7 & 1.3 & 3.2 & 5.2 \\
    &            &            & 37.7 & 5.3 & 7.4 & 25.0 \\
    \midrule
    \multirow{4}{*}{SpT5}
    & \checkmark & \checkmark & 5.6& 1.1 & 1.2 & 3.3\\
    & \checkmark &            & 8.8& 1.9 & 1.7 & 5.2\\
    &            & \checkmark & 7.8 & 1.7 & 1.5 & 4.6 \\
    &            &            & 10.0& 2.4 & 2.2 & 5.4\\
    \bottomrule
  \end{tabularx}
  \caption{Ablation study on encoder, adapter, and A-RoPE. WER and its decomposition into deletion (Del), insertion (Ins), and substitution (Sub) rates are reported. The proposed configuration is shown in bold. GT denotes ground truth recordings. SpT5: SpeechT5.}
  \label{tab:ablation}
\end{table}

We compare the proposed model (ByT5 + Adapter + A-RoPE) against ablation variants to evaluate the contribution of each component. Results are summarized in \autoref{tab:ablation}.
The proposed model achieves a WER of 5.1\%, comparable to that obtained on the ground truth (human-recorded) audio, which reflects the inherent transcription difficulty of rap vocals for ASR systems. 

Removing A-RoPE increases WER only marginally to 6.0\%. Notably, this ByT5 + Adapter configuration alone already outperforms every SpeechT5-based variant except the one combining both the adapter and A-RoPE (5.6\%), suggesting that a lightweight adapter can substitute for costly speech pretraining as a means of achieving effective cross-modal grounding.
Removing the adapter while retaining A-RoPE yields a WER of 9.7\%, on par with SpeechT5 without any adaptation (10.0\%). 
This suggests that A-RoPE's positional prior can substantially compensate for the lack of a learned cross-modal representation, by directly encouraging sequential correspondence between lyric and audio tokens.
Without both adapter and A-RoPE, the model produces significantly degraded lyric coverage, reaching a WER of 37.7\%. These results collectively demonstrate that neither component alone is sufficient to match the full configuration.

Together, the adapter and A-RoPE are complementary: the adapter provides cross-modal grounding while A-RoPE supplies the positional structure needed to maintain lyric coverage under strong beat conditioning.
This complementarity is more clearly observed in the SpeechT5 results, where the modality gap is already considerably addressed through speech pretraining. Adding A-RoPE alone yields a larger improvement (10.0\% $\rightarrow$ 7.8\%) than adding the adapter alone (10.0\% $\rightarrow$ 8.8\%), suggesting that under sufficient cross-modal grounding, A-RoPE's positional alignment prior has a stronger effect on lyric coverage.
The full effect of A-RoPE is further examined on variable tempo settings in \autoref{sec:tempo}.
When both components are combined, SpeechT5 nearly matches the proposed ByT5 configuration. Despite the similar WER, we retain ByT5 for its byte-level flexibility, which avoids the vocabulary limitations of SpeechT5 discussed in \autoref{sec:input_ref}.

\subsection{Training Dynamics}
\begin{figure}[t]
    \centering
    \begin{subfigure}[b]{0.49\columnwidth}
        \centering
        \includegraphics[width=\linewidth]{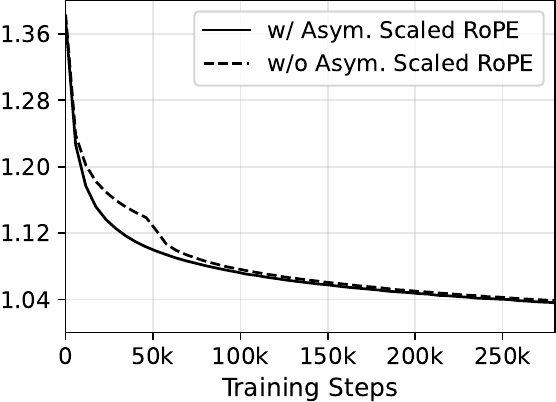}
        \caption{Training loss}
        \label{fig:training_loss}
    \end{subfigure}
    \hfill
    \begin{subfigure}[b]{0.49\columnwidth}
        \centering
        \includegraphics[width=\linewidth]{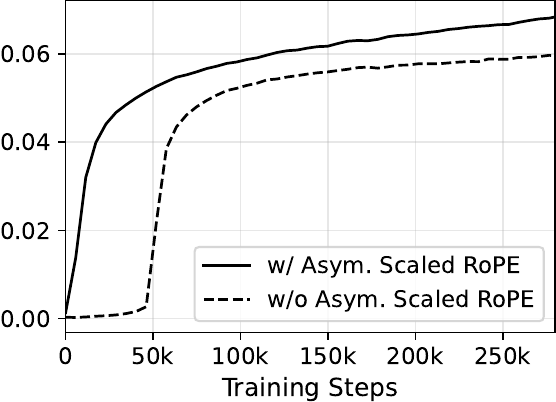}
        \caption{Lyrics utilization}
        \label{fig:training_utilization}
    \end{subfigure}
    \caption{Training curves comparing models with and without A-RoPE. (a) Training loss. (b) Lyrics utilization, measured as the relative loss increase when replacing ground-truth lyrics with irrelevant lyrics in cross-attention.}
    \label{fig:training_curves}
\end{figure}
To quantify how effectively the model utilizes lyric content during training, we define a lyrics utilization score as $(\mathcal{L}_\text{irr} - \mathcal{L}_\text{gt}) / \mathcal{L}_\text{gt}$, where $\mathcal{L}_\text{gt}$ is the loss computed with ground-truth lyrics and $\mathcal{L}_\text{irr}$ is the loss computed with irrelevant lyrics from a different sample in the same batch. This score can be roughly interpreted as the relative contribution of lyric content to the model's loss. 

As shown in \autoref{fig:training_curves}, the model trained with A-RoPE exhibits immediate and steady growth in lyrics utilization from the start of training, whereas the model without A-RoPE shows near-zero utilization for approximately 50k steps before rising sharply. This delayed activation coincides with a visible discontinuity in the training loss curve, suggesting that the model abruptly shifts its reliance toward lyric content once the cross-attention becomes effective. Although both models converge to similar training loss, the following section shows that this does not translate into equivalent generalization under tempo mismatch.

\subsection{Tempo Robustness}\label{sec:tempo}

\begin{table}[t]
  \centering
  \small
  \begin{tabularx}{\columnwidth}{C|P{2.1}P{1.1}P{1.1}|P{2.1}P{1.1}P{1.1}}
    \toprule
    \multirow{2}{*}[-0.5em]{\makecell{Speed \\ (CPS)}} & \multicolumn{3}{c|}{w/ A-RoPE} & \multicolumn{3}{c}{w/o A-RoPE} \\
    \cmidrule(lr){2-4}\cmidrule(l){5-7}
    & \mc{WER\hspace{0.1em}$\downarrow$} & \mc{Del\hspace{0.1em}$\downarrow$} & \mcc{Ins\hspace{0.1em}$\downarrow$} & \mc{WER\hspace{0.1em}$\downarrow$} & \mc{Del\hspace{0.1em}$\downarrow$} & \mc{Ins\hspace{0.1em}$\downarrow$} \\
    \midrule
    $-2\sigma$ & 20.5 & 2.2& 9.3& 33.6& 5.7& 14.6\\
    $-1\sigma$ & 14.2& 2.6& 5.2& 25.5& 5.9& 8.3\\
    $\mu$ & 12.3 & 3.3& 2.8& 22.0& 7.2& 4.7\\
    $+1\sigma$ & 13.8& 4.1& 2.2& 24.4& 9.3& 3.9\\
    $+2\sigma$ & 15.4& 5.8& 2.2& 26.9& 12.6& 3.1\\
    \bottomrule
  \end{tabularx}
\caption{Tempo robustness evaluation (WER/Del/Ins, \%) for the proposed model and its ablation without A-RoPE. Lyrics and beat tracks are randomly mismatched across test samples. Speed bin labels denote the offset from the training mean $\mu$, each spanning 1$\sigma$ ($\mu$=17.9, $\sigma$=2.9 chars/sec).}
  \label{tab:tempo}
\end{table}

To evaluate tempo robustness, we randomly mismatch lyrics and beat tracks from the test set, controlling the ratio of lyric length to beat duration. Specifically, we partition the mismatched pairs into five bins based on the characters-per-second (CPS) rate, using the training distribution ($\mu=17.9$, $\sigma=2.9$ chars/sec) to define bin boundaries.
Each bin spans $1\sigma$ and is centered at $\mu-2\sigma$ (slowest), $\mu-1\sigma$, $\mu$, $\mu+1\sigma$, and $\mu+2\sigma$ (fastest).

As shown in \autoref{tab:tempo}, the proposed model consistently achieves lower WER across all bins, with the largest improvements observed in the extreme bins where lyrics are significantly faster or slower than the beat.
We note that WER values in this experiment are higher than those reported in \autoref{tab:ablation}, as the lyrics are not written to match the given beat. In the matched setting, lyrics are composed to fit the rhythmic structure of the beat, making it easier for the model to distribute syllables naturally. In the mismatched setting, the model must adapt its pacing to accommodate lyrics of varying lengths relative to the beat, which inherently increases the likelihood of deletion and insertion errors. The error pattern is directional: deletions increase at higher CPS while insertions increase at lower CPS, consistent with the skip and repeat behaviors described in \autoref{sec:rope}.
 
\subsection{Speaker Similarity and Audio Quality}

\begin{table}[t]
 \small
 \centering
 \begin{tabularx}{\columnwidth}{@{\hspace{3pt}}c@{\hspace{5pt}}cCC}
    \toprule
    Model & Group & \mbox{WavLM Sim $\uparrow$} & \mbox{DNSMOS OVR $\uparrow$} \\
    \midrule
    GT & --- & 0.80 $\pm$ 0.05 & 3.35 $\pm$ 0.21 \\
    \cmidrule(lr){1-4}
    \multirow{2}{*}{w/ A-RoPE}
    & \textit{match} & 0.80 $\pm$ 0.05 & 3.49 $\pm$ 0.20 \\
    & \textit{mismatch} & 0.75 $\pm$ 0.05 & 3.46 $\pm$ 0.21 \\
    \cmidrule(lr){1-4}
    \multirow{2}{*}{w/o A-RoPE}
    & \textit{match} & 0.80 $\pm$ 0.05 & 3.46 $\pm$ 0.20 \\
    & \textit{mismatch} & 0.76 $\pm$ 0.05 & 3.42 $\pm$ 0.22 \\
    \bottomrule
  \end{tabularx}
  \caption{Speaker similarity (WavLM Sim) and audio quality (DNSMOS OVR) under matched and mismatched conditions (mean $\pm$ std).}
  \label{tab:similarity}
\end{table}

\autoref{tab:similarity} reports speaker similarity and audio quality under matched (original lyric-beat pairs) and mismatched (randomly reassigned pairs from \autoref{sec:tempo}) conditions. In the matched setting, both models achieve a speaker similarity of 0.80, matching the GT upper bound and demonstrating effective speaker identity preservation from a 3-second reference voice. Under mismatched conditions, similarity decreases slightly to 0.75--0.76, likely due to the increased difficulty of maintaining consistent vocal delivery when lyrics and beat are not originally paired. The difference between the two models remains within the standard deviation, indicating that A-RoPE does not affect timbre reproduction.

DNSMOS scores are consistent across all conditions, with both models producing audio quality comparable to or slightly above GT (3.35). The higher DNSMOS of generated vocals (3.42--3.49) can be attributed to the smoothing effect of the generation process, which tends to suppress background noise and vocal artifacts present in real recordings. Under mismatched conditions, the model with A-RoPE scores marginally higher than the variant without it (3.46 vs.\ 3.42), indicating that improved lyric alignment does not come at the cost of audio quality.

\subsection{Subjective Evaluation}\label{sec:human_mos}

We conduct a listening study to evaluate rhythmic flow (MOS-F), speaker similarity (MOS-S), and audio quality (MOS-Q) on a 5-point Likert scale. Two model variants are evaluated: the proposed model and its ablation without A-RoPE. For each model, 15 clips are randomly sampled from the generated outputs of the mismatched test set (3 per CPS bin), selecting only samples for which both models achieve WER below 5\%. Without this control, differences in lyric accuracy dominate listener judgments: listeners tend to penalize samples with audible mispronunciations or missing words, overshadowing the perceptual dimensions we aim to evaluate. 
For speaker similarity, listeners are provided with the 3-second reference voice that was used as input during generation.
Additionally, 10 ground truth clips are randomly sampled from the matched test set, with the reference voice drawn from a different segment by the same rapper. All 40 clips are randomly shuffled and presented to 40 listeners in a blind evaluation without any indication of their source.

\begin{table}[t]
  \centering
  \small
    \begin{tabularx}{\columnwidth}{c|CCC}
      \toprule
      Model & MOS-F $\uparrow$ & MOS-S $\uparrow$ & MOS-Q $\uparrow$\\
      \midrule
      GT & 3.81 & 3.95 & 3.73\\
      w/ A-RoPE & 3.51 & 3.82 & 3.35 \\
      w/o A-RoPE & 3.34  & 3.98  & 3.46 \\
      \bottomrule
    \end{tabularx}
  \caption{Subjective evaluation results (MOS, 5-point scale). MOS-F: rhythmic flow, MOS-S: speaker similarity, MOS-Q: audio quality.}
  \label{tab:mos}
\end{table}

Results are summarized in \autoref{tab:mos}. Statistical significance is assessed using Welch's t-test, which does not assume equal variances across groups. 
The proposed model achieves significantly higher MOS-F than the ablation ($p$ = 0.01), suggesting that A-RoPE contributes to improved perceived rhythmic flow. No significant difference was found in MOS-Q ($p$ = 0.06).
Both models score below GT in flow and audio quality ($p$ < 0.001), which is expected, as GT clips reflect naturally aligned lyric-beat pairs, whereas the models are evaluated under mismatched conditions.
For speaker similarity, neither model differs significantly from GT ($p$ > 0.05), indicating that both models preserve speaker identity at a level comparable to ground truth recordings.

\section{Conclusion}
 
We presented BARS, a Flow Matching-based rapping voice generation model, conditioned on a beat track, lyrics, and a reference voice. By introducing A-RoPE in cross-attention, the model achieves robust lyric-audio alignment without explicit duration prediction or phoneme-level supervision, substantially reducing lyric skipping and repetition under varying tempo conditions. A lightweight Transformer adapter bridges the modality gap between text-only pretrained ByT5 embeddings and audio latents, matching or exceeding the performance of speech-pretrained text encoders while retaining byte-level flexibility.
Experiments demonstrate that the proposed model achieves near-ground-truth intelligibility on matched lyric-beat pairs, and maintains strong performance under mismatched conditions where the vocal pacing deviates significantly from the training distribution. Ablation studies confirm that both the A-RoPE mechanism and the adapter are essential to the model's effectiveness.

A current limitation is that our evaluation is restricted to English. While ByT5 operates at the byte level and is inherently language-agnostic, differences in byte-per-character ratios across non-ASCII languages may affect the relationship between lyric length and audio duration. Our demo page includes preliminary multilingual examples, but systematic evaluation and adaptation of the scaling mechanism are left to future work.

\section{Ethics Statement}
The proposed model, BARS, is capable of zero-shot timbre reproduction from a short reference voice clip. While this functionality is designed as a creative tool for rap artists to prototype and refine their vocal delivery, we acknowledge the potential for misuse, such as the unauthorized synthesis of a specific individual's voice. We emphasize that this technology should be used responsibly and only with the explicit consent of the original speaker.

\section{AI Usage Statement} No AI tools were used as part of the core methodology described in this paper. AI-assisted writing tools were used solely for editing purposes, including spell checkers and grammar correction. All content was reviewed and verified by the authors.

\bibliography{ISMIRtemplate}

\end{document}